\documentclass{jfm}
\usepackage{graphicx}
\usepackage{epstopdf, epsfig}

\usepackage{subcaption}
\usepackage{hyperref}% add hypertext capabilities
\usepackage{color}% add color
\newcommand{\aleck}{\textcolor{black}}
\newcommand{\bernhard}{\textcolor{black}}
\newcommand{\Rone}{\textcolor{black}}
\newcommand{\Rtwo}{\textcolor{black}}
\definecolor{orange}{rgb}{1,0.5,0}
\newcommand{\Rthree}{\textcolor{black}}

\shorttitle{An efficient cellular flow model for cohesive particle flocculation in turbulence}
\shortauthor{K. Zhao, B. Vowinckel, T.-J. Hsu, T. K\"{o}llner, B. Bai and E. Meiburg}

\title{An efficient cellular flow model for cohesive particle flocculation in turbulence}

\author{K. Zhao\aff{1,2},
  B. Vowinckel\aff{1,3},
  T.-J. Hsu\aff{4},
  T. K\"{o}llner\aff{1,5},
  B. Bai\aff{2},
 \and E. Meiburg\aff{1}\corresp{\email{meiburg@engineering.ucsb.edu}}
        }

\affiliation{\aff{1}Department of Mechanical Engineering, UC Santa Barbara, CA 93106, USA
\aff{2}State Key Laboratory of Multiphase Flow in Power Engineering, Xi'an Jiaotong University, Xi'an 710049, China
\aff{3}Leichtwei\ss-Institut f\"{u}r Wasserbau, Technische Universit\"{a}t Braunschweig, 38106 Braunschweig, Germany
\aff{4}Center for Applied Coastal Research, Department of Civil \& Environmental Engineering, University of Delaware, Newark, DE 19716, USA
\aff{5}Present address: CADFEM GmbH, 85567 Grafing, Germany
}

\begin{document}

\maketitle

\begin{abstract}
%We explore the flocculation of cohesive sediment in steady cellular Taylor-Green vortices via one-way coupled simulations. 
\Rone{
We propose a one-way coupled model that tracks individual primary particles in a conceptually simple cellular flow setup to predict flocculation in turbulence. This computationally efficient model accounts}
 for Stokes drag, lubrication, cohesive and direct contact forces \Rone{on the primary spherical particles}, and 
 \Rone{ allows for a systematic simulation campaign that yields} 
 the transient mean floc size as a function of the governing dimensionless parameters. 
 \Rone{ The simulations}
 reproduce the growth of the cohesive flocs with time, and the emergence of a log-normal equilibrium distribution governed by the balance of aggregation and breakage. Flocculation proceeds most rapidly when the Stokes number of the primary particles is \textit{O}(1). Results from this simple computational model
are consistent with experimental observations, thus allowing us to propose a new analytical flocculation model that yields improved agreement with experimental data, especially during the transient stages. 
\end{abstract}

\begin{center}
------------------------------------------------------------------------------------------------------------------
\end{center}

%\begin{keywords}
%Authors should not enter keywords on the manuscript, as these must be chosen by the author during the online submission process and will then be added during the typesetting process (see http://journals.cambridge.org/data/\linebreak[3]relatedlink/jfm-\linebreak[3]keywords.pdf for the full list)
%\end{keywords}

\section{Introduction}
Cohesive sediment, commonly defined as particles with diameters $D_p < 63 \mathrm{\mu m}$, plays a central role in a wide range of environmental and industrial processes. 
%\citep{Grabowski2011} 
For these small grain sizes, attractive van der Waals forces can outweigh hydrodynamic, buoyancy and collision forces, and trigger the formation of large aggregates via flocculation \citep{Yashimasa2017}. Following the pioneering work by \citet{Levich1962floc}, current approaches for modeling the flocculation process often employ population balance equations \citep{Maggi2007,  Verney2011, Shin2015}
\bernhard{or simplified versions thereof}
%. Simplified versions of this approach have been developed as well 
\citep{Winterwerp1998floc, Son2008floc, Son2009floc, Lee2011, Shen2018}. These semi-empirical models, which require calibration with experimental data, usually do not account for the detailed profiles of the various forces governing particle-particle interactions.

The present investigation 
\bernhard{presents a conceptually simple model to obtain flocculation data}
%aims to gain further physical insight into the mechanisms governing flocculation 
via one-way coupled \Rone{simulations that track individual primary particles and} accurately capture the inter-particle forces, based on the recent development of advanced collision models in viscous flows \citep[][and references therein]{Biegert2017Collision}, along with strategies for accurately modeling cohesive forces \citep{Vowinckel2018Cohesive}. Towards this end, we employ the well known initial configuration of cellular Taylor-Green flow as a simple, quasi-steady analytical model of a turbulent flow at the Kolmogorov scale. This flow has previously been used successfully in elucidating fundamental aspects of particle-vortex interactions \citep{Maxey1987Cellular, Guazzelli2014Cellular}. 
\Rone{We will exploit this conceptually simple, computationally efficient scenario to systematically investigate the influence of key physical parameters, and propose a new flocculation model that agrees closely with experimental data.}

\begin{figure}
    \begin{subfigure}{0.5\textwidth}
    \centering
    \caption{}
    \includegraphics[width=0.7\textwidth]{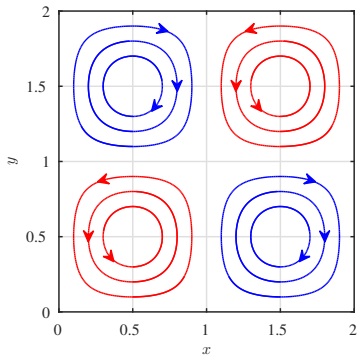}
    \label{fig:streamlines}
    \end{subfigure}
    \begin{subfigure}{0.5\textwidth}
    \centering
    \caption{}
    \includegraphics[width=0.7\textwidth]{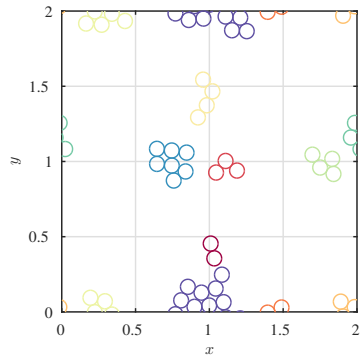}
    \label{fig:floc configuration}
    \end{subfigure}
    \caption{(a) Streamlines of the doubly periodic
    %idealized 
    background flow; (b) Typical floc configuration \Rone{made up of spherical primary particles}, with individual flocs distinguished by color.}
\end{figure}

\section{Computational model}\label{sec:Computational model}
\subsection{Particle motion in cellular flow fields} \label{subsec:Particle motion in cellular flow fields}
In the spirit of earlier investigations by \citet{Maxey1987Cellular} and \citet{Guazzelli2014Cellular}, we apply a simple model flow in order to investigate the effects of turbulence on the dynamics of cohesive particles. 
%More specifically, 
\bernhard{We}
consider the one-way coupled motion of small spherical particles in the two-dimensional, steady, spatially periodic cellular flow field commonly employed as initial condition for simulating Taylor-Green vortices (cf. figure \ref{fig:streamlines}), with fluid velocity field ${\boldsymbol u}_f = (u_f, v_f)^{\rm T}$
\begin{equation}
    u_f = \frac {U_0}{\pi} {\rm sin}\left(\frac{\pi x}{L}\right) {\rm cos}\left( \frac{\pi y}{L}\right) \\\\\\\ , \\\\\\\ v_f = -\frac {U_0}{\pi} {\rm cos}\left( \frac{\pi x}{L}\right) {\rm sin}\left( \frac{\pi y}{L}\right) \ ,
\end{equation} 
where $L$ and $U_0$ represent the characteristic length and velocity scales of the vortex flow. 

\Rone{Keeping in mind that cohesive sediment grains in nature may be non-spherical, we nevertheless approximate each primary particle $i$ as a sphere that} moves with the translational velocity ${\boldsymbol u}_{p,i} = (u_{p,i}, v_{p,i})^{\rm T}$ and the angular velocity $\omega_{p,i}$. These are obtained from the linear and angular momentum equations
\begin{equation} \label{eq:momentum}
    m_p \frac{\mathrm{d}{\boldsymbol u}_{p,i}}{\mathrm{d}t} = {\boldsymbol F}_{d,i} + {\boldsymbol F}_{g,i} + \underbrace{\sum_{j=1,j \ne i}^{N_p}(\boldsymbol F_{con,ij} + \boldsymbol F_{lub,ij} + \boldsymbol F_{coh,ij})}_{{\boldsymbol F}_{c,i}} \ ,
\end{equation}
\begin{equation}
    I_p \frac{\mathrm{d}{\boldsymbol \omega_{p,i}}}{\mathrm{d}t} = \underbrace{\sum_{j=1,j \ne i}^{N_p}(\boldsymbol T_{con,ij} + \boldsymbol T_{lub,ij})}_{{\boldsymbol T}_{c,i}} \ ,
\end{equation}
where \Rone{the primary particle $i$} moves in response to the Stokes drag force $\boldsymbol F_{d,i} = -3 \pi D_p \mu_f ({\boldsymbol u}_{p,i} - {\boldsymbol u}_{f,i})$, the gravitational force $\boldsymbol F_{g,i} = \pi D_p^3 (\rho_p - \rho_f)\boldsymbol g / 6$, and the particle-particle interaction force $\boldsymbol F_{c,i}$. Here ${\boldsymbol u}_{f,i}$ and ${\boldsymbol u}_{p,i}$ indicate the fluid and particle velocities evaluated at the particle center. $m_p$ denotes the particle's mass, $D_p$ its diameter, $\rho_p$ its density, and $N_p$ the total number of particles in the flow. We assume all particles to have the same diameter and density. $\mu_f$ and $\rho_f$ denote the dynamic viscosity and the density of the fluid, respectively, and $\boldsymbol g$ is the gravitational acceleration. $\boldsymbol F_{c,i}$ accounts for the direct contact force $\boldsymbol F_{con,ij}$ in normal and tangential direction, as well as for short-range forces due to lubrication $\boldsymbol F_{lub,ij}$ and cohesion $\boldsymbol F_{coh,ij}$, where the subscript $ij$ indicates the interaction between particles $i$ and $j$. $I_p = \pi \rho_p D_p^5 / 60$ denotes the moment of inertia of a particle. $\boldsymbol T_{c,i}$ represents the torque due to particle-particle interactions, where we distinguish between the direct contact torque $\boldsymbol T_{con,ij}$ and lubrication torque $\boldsymbol T_{lub,ij}$.

Following \citet{Biegert2017Collision}, we represent the direct contact force $\boldsymbol F_{con,ij}$ by means of spring-dashpot functions, while the lubrication force $\boldsymbol F_{lub,ij}$ is accounted for based on \citet{Cox1967I} as implemented in \citet{Biegert2017PEPS}.
%Implementation details are provided in \citet{Biegert2017PEPS}. 
The model for the cohesive force $\boldsymbol F_{coh,ij}$ is based on the work of \citet{Vowinckel2018Cohesive}. It assumes a parabolic force profile, distributed over a thin shell surrounding each particle. 

\subsection{Non-dimensionalization} \label{subsec:Non-dimensionalization}
We choose $L$, $U_0$ and $L/U_0$ as the characteristic length, velocity and time scales. Conceptually, these can be thought of as representing Kolmogorov scales. In this way, we obtain the dimensionless equation of motion for the particles as 
\begin{equation} \label{eq:dimensionless momentum}
    \tilde m_p \frac{\mathrm{d} \tilde {\boldsymbol u}_{p,i}}{\mathrm{d} \tilde t}  =  \underbrace{- \frac {\tilde m_p (\tilde {\boldsymbol u}_{p,i} - \tilde {\boldsymbol u}_{f,i})}{St}}_{\tilde {\boldsymbol F}_{d,i}} + \underbrace{\frac{\tilde m_p \tilde W}{St}}_{\tilde {\boldsymbol F}_{g,i}} + \sum_{j=1,j \ne i}^{N_p}(\tilde {\boldsymbol F}_{con,ij} + \tilde {\boldsymbol F}_{lub,ij} + \tilde {\boldsymbol F}_{coh,ij}) \ ,
\end{equation}
where dimensionless quantities are denoted by a tilde. The dynamics of the 
\Rone{primary particles}
are characterized by the Stokes number $St = U_0 \rho_p D_p^2 / (18 L \mu_f)$ and the settling velocity $\tilde W = v_s / U_0$, where $v_s = (\rho_p - \rho_f) D_p^2 \boldsymbol g/ (18 \mu_f)$ is the Stokes settling velocity 
\Rone{of an individual, isolated primary particle.}
The dimensionless particle mass and density ratio are defined as $\tilde m_p = \pi \tilde D_p^3 \tilde \rho_s/6$ and $\tilde \rho_s = \rho_p / \rho_f$, respectively.

The dimensionless direct contact force $\tilde {\boldsymbol F}_{con,ij}$ between particles includes the normal component $\tilde {\boldsymbol F}_{con,n,ij}$ and the tangential component $\tilde {\boldsymbol F}_{con,t,ij}$, which are defined as
\begin{equation}
    \tilde {\boldsymbol F}_{con,ij} = \left\{
    \begin{array}{ll}
        \underbrace{- \tilde k_n \vert{\tilde \zeta_{n, ij} - \tilde \zeta_{min}}\vert^{3/2} \boldsymbol n - \tilde d_n \tilde {\boldsymbol v}_{n,ij}}_{\tilde {\boldsymbol F}_{con,n,ij}} + \\ \underbrace{{\rm min} (- \tilde k_t \tilde \zeta_{t,ij} - \tilde d_t \tilde {\boldsymbol v}_{t,ij}, \vert\vert{f \tilde {\boldsymbol F}_{con,n,ij}}\vert\vert )\boldsymbol t}_{\tilde {\boldsymbol F}_{con,t,ij}},          &  \tilde \zeta_{n, ij} \leqslant \tilde \zeta_{min} \ , \\ 
        0,   & \rm{otherwise} \ ,
    \end{array}\right.
\end{equation}
where $\tilde \zeta_{n, ij}$ is the normal surface distance between particles $i$ and $j$. 
%In order to achieve a more realistic collision, 
We account for the surface roughness of the particles, which is set to $\tilde \zeta_{min} = 0.0015 \tilde D_p$. $\tilde \zeta_{t, ij}$ is the tangential spring displacement, which denotes the accumulated relative tangential motion between two particles in contact. $\tilde {\boldsymbol v}_{n,ij}$ and $\tilde {\boldsymbol v}_{t,ij}$ denote the normal and tangential components of the relative velocity of particles $i$ and $j$. $\boldsymbol n$ represents the outward-pointing normal on the particle surface, and $\boldsymbol t$ points in the direction of the tangential force. 
\Rthree{We use the parametrization for silicate grains described in \citet{Biegert2017Collision}, so that we chose a standard friction coefficient of $f=0.15$ and obtain stiffness $\tilde k_n$ and $\tilde k_t$ and damping $\tilde d_n$ and $\tilde d_t$ to obtain a specified restitution coefficient as the ratio of impact to rebound velocity ($e_n=0.97$) for the normal component and rolling conditions for the tangential component of $\tilde {\boldsymbol F}_{con,ij}$, respectively.}

The dimensionless lubrication force $\tilde {\boldsymbol F}_{lub,ij}$ between particles $i$ and $j$ has the normal and tangential components $\tilde {\boldsymbol F}_{lub,n,ij}$ and $\tilde {\boldsymbol F}_{lub,t,ij}$, respectively, which are defined as
\begin{equation}
    \tilde {\boldsymbol F}_{lub,ij} = \left\{
    \begin{array}{ll}
        \underbrace{-\frac{\tilde m_p \tilde D_p \tilde {\boldsymbol v}_{n,ij}}{8 St \tilde \zeta_{n,ij}}}_{\tilde {\boldsymbol F}_{lub,n,ij}} + \underbrace{\frac{\tilde m_p}{2 St} (k_1 \tilde {\boldsymbol u}_{t,ij} + k_2 \tilde {\boldsymbol w}_{t,ij})}_{\tilde {\boldsymbol F}_{lub,t,ij}},        &  \tilde \zeta_{min} < \tilde \zeta_{n,ij} \leqslant \tilde h \ , \\ 
        0,   & \rm{otherwise} \ ,
    \end{array}\right.
\end{equation}
where $\tilde h = \tilde D_p/10$ is the range of the lubrication force, $\tilde {\boldsymbol u}_{t,ij}$ and $\tilde {\boldsymbol w}_{t,ij}$ denote the tangential components of the relative translational velocity and the relative rotational velocity of the particles, respectively. The coefficients $k_1$ and $k_2$ take the values $k_1 = 0.53\ln (4 \tilde \zeta_{n,ij} / \tilde D_p) - 0.9588$ and $k_2 = 0.13\ln (4 \tilde \zeta_{n,ij} / \tilde D_p) - 0.2526$, respectively \citep{Biegert2017PEPS}.

The dimensionless cohesive force $\tilde {\boldsymbol F}_{coh,ij}$ is defined as
\begin{equation}
    \tilde {\boldsymbol F}_{coh,ij} = \left\{
    \begin{array}{ll}
        - 4 Co \frac {\tilde \zeta_{n,ij}^2 - \tilde \lambda \tilde \zeta_{n,ij}}{\tilde \lambda^2} \boldsymbol n,        &  \tilde \zeta_{min} < \tilde \zeta_{n,ij} \leqslant \tilde \lambda \ , \\ 
        0,   & \rm{otherwise} \ ,
    \end{array}\right.
\end{equation}
where $\tilde \lambda = \tilde h /2 = \tilde D_p/20$ represents the range of the cohesive force. The cohesive number $Co$ indicates the ratio of the maximum cohesive force $\vert\vert{\boldsymbol F_{coh,ij}}\vert\vert$ at $\tilde \zeta_{n,ij} = \tilde \lambda/2$ to the characteristic inertial force
\begin{equation}
    Co = \frac {{\rm max} (\vert\vert{\boldsymbol F_{coh,ij}}\vert\vert)}{U_0^2 L^2 \rho_f} = \frac {A_H D_p}{16 \lambda \zeta_{0} } \frac {1}{U_0^2 L^2 \rho_f } \ ,
\end{equation}
where the Hamaker constant $A_H$ is a function of the particle and fluid properties 
%\citep{Hamaker1937Cohesive}, 
and the characteristic distance $\zeta_0  = 0.00025D_p$. \citet{Vowinckel2018Cohesive} provide representative values of $A_H$ for common natural systems.

% In order to better describe the particle motion in vertex, we introduce two important parameters, the Stokes number $St$ which is determined by $Re$, $\tilde \rho_s$ and $\tilde D_p$, the velocity ratio $\tilde W$ which is the ratio of the particle settling velocity $v_s$ to the vortex velocity $U_0$,
% \begin{equation} \label{eq:St and W}
%     St = \frac {1}{18} Re \tilde D_p^2 \tilde \rho_s \ \ \ \ , \ \ \ \    \tilde W = St (1- \frac{1}{\tilde \rho_s}) \tilde {\boldsymbol g}
% \end{equation}
% where $St$ and $\tilde W$ characterise the behavior of particles suspended and settling in vortices, respectively. The particle settling velocity is defined with Stokes drag law, $v_s = (\rho_p - \rho_f) g D_p^2 / 18 \mu_f$. Therefore, the dimensionless drag and gravitational forces in eqn. (\ref{eq:dimensionless momentum}) can be also written as $\tilde {\boldsymbol F}_{d,i} = \tilde m_p(\tilde {\boldsymbol u}_{f,i} - \tilde {\boldsymbol u}_{p,i})/ St$ and $\tilde {\boldsymbol F}_{g,i} = \tilde m_p \tilde W/ St$.
To summarize, the simulations require as input parameters the 
\Rtwo{dimensionless} 
particle diameter $\tilde D_p$, the number of particles $N_p$, the density ratio $\tilde \rho_s$, the settling velocity $\tilde W$, the Stokes number $St$ and the cohesive number $Co$.  For convenience, the tilde symbol will be omitted henceforth.
% \eckart{(EM: if we specify St and W, do we really still need to specify $\rho_s$? I don't think it enters independently of St and W, correct?)}\aleck{(I think $\tilde \rho_s$ should be there. With the fixed $St$, $\tilde W$ and $\tilde D_p$, $Re$ and $\tilde \rho_s$ determine $St$, $\tilde g$ and $\tilde \rho_s$ determine $\tilde W$, so we should have a fixed $\tilde \rho_s$ to make the inputs unique.)}

\subsection{Validation: Aggregation and breakage of two particles}\label{subsec:Aggregation and breakage of two particles}
\begin{figure}
    \begin{subfigure}{0.5\textwidth}
    \centering
    \caption{}
    \includegraphics[width=0.9\textwidth]{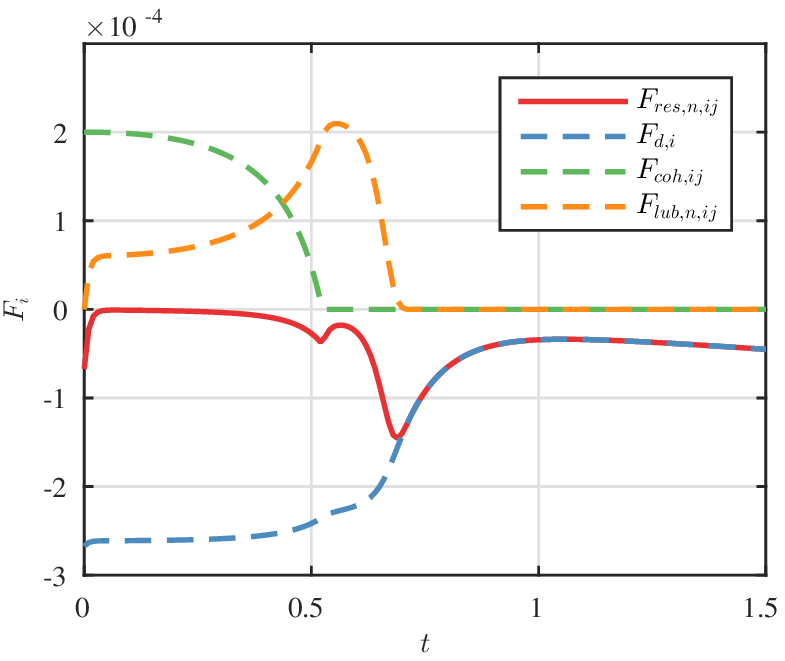}
    \label{fig:binary horizontal force breakage}
    \end{subfigure}
    \begin{subfigure}{0.5\textwidth}
    \centering
    \caption{}
    \includegraphics[width=0.9\textwidth]{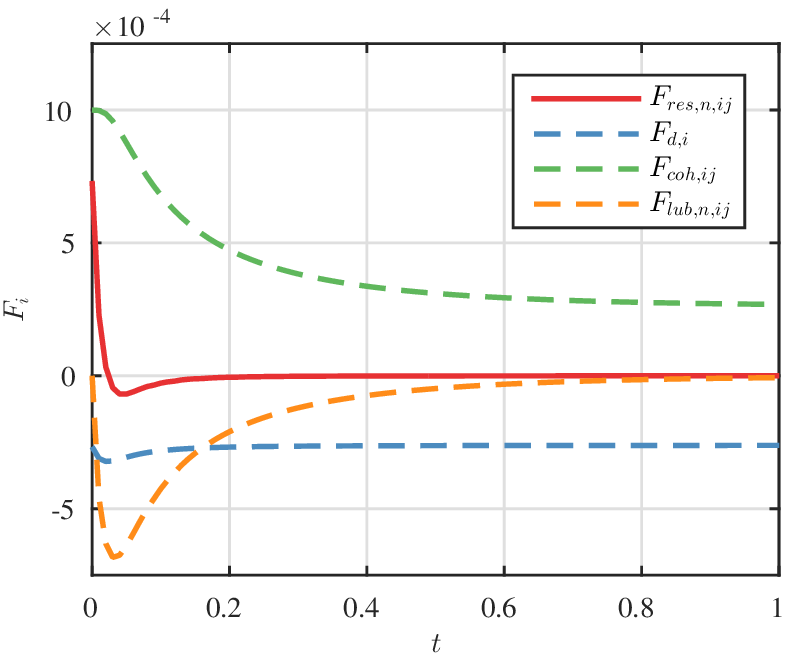}
    \label{fig:binary horizontal force aggregation}
    \end{subfigure}
    \caption{Transient forces on the particle to the left of the stagnation point at (1,1) in figure \ref{fig:streamlines}, during binary interaction. The simulation parameters are $D_p = 0.1$, $\rho_s = 1$, $St = 0.1$, $W = 0$: (a) breakage, $Co = 2 \times 10^{-4}$; (b) aggregation, $Co = 1 \times 10^{-3}$.}
    \label{fig:binary horizontal force}
\end{figure}

To validate our numerical implementation of the cohesive force model, we consider the interaction of two neutrally buoyant particles with $D_p=0.1$, $St=0.1$, $W=0$ that are placed symmetrically to the left and right of the stagnation point at (1,1) in figure \ref{fig:streamlines}. The particles are at rest initially, at a surface distance of $\lambda /2 = 0.0025$, so that the cohesive force is at its maximum. Figure \ref{fig:binary horizontal force} presents the temporal evolution of the various forces acting on the particle to the left of the stagnation point, for the two scenarios of (a) floc breakage, and (b) floc aggregation. 
%In figure \ref{fig:binary horizontal force breakage} we see that 
For the smaller value of $Co$, the drag force that tries to separate the particles is initially larger than the cohesive force that attracts them to each other (figure \ref{fig:binary horizontal force breakage}). As the surface distance between the particles increases, the cohesive force decays and approaches zero. While the lubrication force $F_{lub,n,ij}$ acts to slow the separation of the particles, the overall net force $F_{res,n,ij}$ acting on the particle is always negative, so that the particles gradually move apart. When the surface distance between the particles becomes larger than the range of the cohesive and lubrication forces, the net force equals the drag force.

Figure \ref{fig:binary horizontal force aggregation}, on the other hand, focuses on a case in which the cohesive force initially is larger than the drag force, so that the particles approach each other. This process is slowed down by the lubrication force. The particles asymptotically approach an equilibrium position of near contact in which the separating drag force is balanced by the attractive cohesive force.

\section{Large ensemble of particles}\label{sec:Settling a large ensemble}
\subsection{Computational setup} \label{subsec:Typical computational condition}

We now investigate ensembles involving more particles, to obtain insight into the flocculation dynamics of larger systems. %Towards that end,
We employ a computational domain of size $L_x \times L_y = 2 \times 2$, with periodic boundaries (figure \ref{fig:streamlines}). All particles have identical diameters and densities. Initially they are at rest and separated, and randomly distributed throughout the domain. When the distance between two particles is less than $\lambda /2$, we consider them as part of the same floc. We then track the number of flocs $N_f$ as a function of time, with an individual particle representing the smallest possible floc. To improve the statistics, we repeat each simulation five times for different random initial conditions
\Rthree{as the simulation results are statistically independent of the initial particle placement.}

\begin{figure}
    \begin{subfigure}{0.5\textwidth}
    \centering
    \caption{}
    \includegraphics[width=0.9\textwidth]{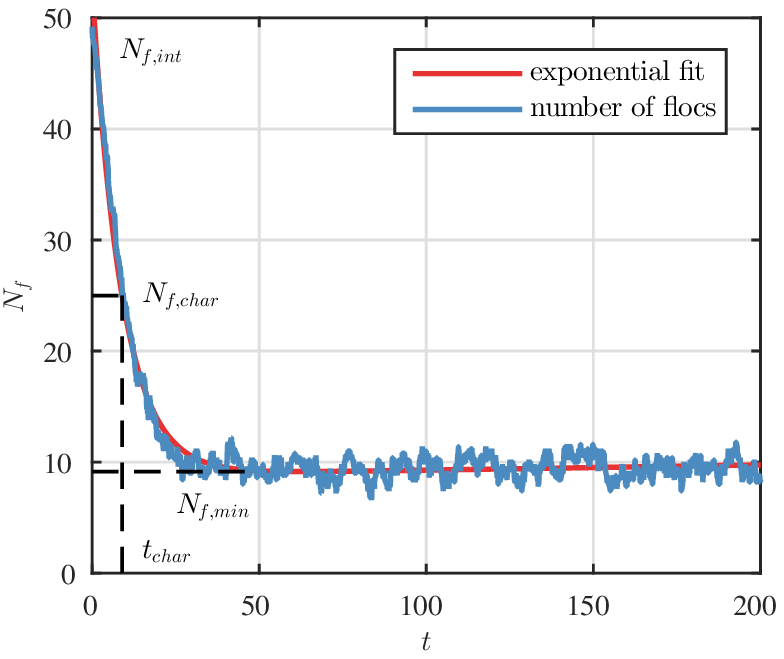}
    \label{fig:particles50 typical Nf}
    \end{subfigure}
    \begin{subfigure}{0.5\textwidth}
    \centering
    \caption{}
    \includegraphics[width=0.9\textwidth]{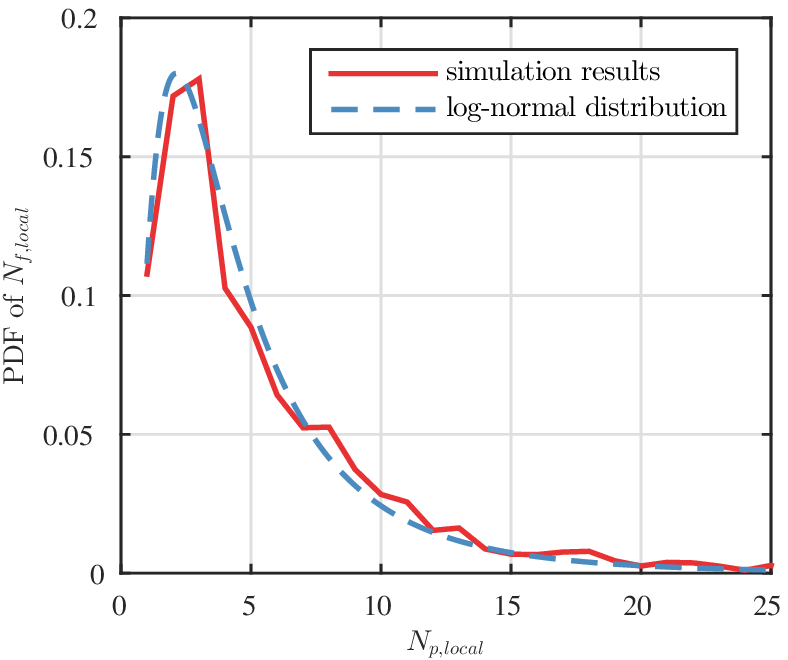}
    \label{fig:particles50 typical PDF}
    \end{subfigure}
    \caption{(a) Typical evolution of the number of flocs $N_f$ as function of time; (b) Floc size distribution during the equilibrium stage $50 \leqslant t \leqslant 200$. Simulation parameters are $N_p = 50$, $D_p = 0.1$, $\rho_s = 1$, $W = 0$, $St = 0.1$, and $Co = 5 \times 10^{-4}$.}
    \label{fig:particles50 typical}
\end{figure}

A typical floc configuration is shown in figure \ref{fig:floc configuration}. Figure \ref{fig:particles50 typical Nf} presents results for a series of simulations with $N_p = 50$ particles 
\Rtwo{that have a size of 10\% of the Kolomogorov length scale, i.e. $D_p = 0.1$. Further, the parameters for this scenario were $\rho_s = 1$, $W = 0$, $St = 0.1$, and $Co = 5 \times 10^{-4}$.}
Since the particles are dispersed initially, the initial number of flocs $N_{f,int} \approx N_p$. Subsequently $N_f$ decreases rapidly due to flocculation, before leveling off around a constant value $N_{f,min}$ that reflects a stable balance between aggregation and breakage. The transient variation of $N_f(t)$ can be fitted by an exponential function of the form
\begin{equation} \label{eq:exponential fitting}
         N_f = (N_{f,int} - N_{f,min}) e^{b t} +  N_{f,min} \ .
\end{equation}
where we evaluate $N_{f,min}$ as the average number of flocs during the equilibrium stage $50 \leqslant t \leqslant 200 $. 
\Rone{The agglomeration rate $\left|b\right|$ with the constraint $b\leq 0$ is obtained via a least-square fit.
We define the characteristic flocculation time scale $t_{char}$ as the time it takes for the number of flocs to decrease from its initial value $N_{f,int}$ to a characteristic number of flocs $N_{f,char} = N_p /2$. Hence  the corresponding characteristic time can be calculated as $t_{char} = {\rm ln} [(N_p /2 - N_{f,min}) / (N_{f,int} - N_{f,min})] / b$.}
% \begin{equation} 
%          t_{char} = \frac{1}{b} {\rm ln} \left( \frac{N_p /2 - N_{f,min}}{N_{f,int} - N_{f,min}} \right)
% \end{equation}

Figure \ref{fig:particles50 typical PDF} displays the statistical floc size distribution during the equilibrium stage $50 \leqslant t \leqslant 200$, where the "floc size" $N_{p,local}$ denotes the number of particles in a floc. $N_{f,local}$ refers to the number of the flocs of the same size. We find that the floc size distribution closely follows a log-normal distribution, consistent with previous experimental observations \citep{Bouyer2004, Verney2011, Hill2011}.

\subsection{Influence of the governing parameters on the flocculation dynamics} 
\label{subsec:Dependence of flocculation on the governing parameters}

\begin{table}
  \begin{center}
\def~{\hphantom{0}}
  \begin{tabular}{ccccccc}
    Parameter & $Co$ & $St$ & $D_p$ & $\phi$ & $\rho_s$ & $W$\\ 
    \\
    Range \ \ & 0.00015 - 0.025 \ \ & 0.01 - 9 \ \ & 0.04 - 0.1 \ \ & 0.0042 - 0.0916 \ \ & 1 - 3 \ \ & 0 - 2\\ 
    \\
  \end{tabular}
  \caption{Simulated parameter ranges. The particle number $N_p$ is converted into the pseudo volume fraction $\phi = (\pi N_p D_p^3 ) / (6 L_x L_y D_p)$. 
  \Rtwo{Note that $D_p$ is a dimensionless value normalized by the characteristic length scale $L$.}
  }
  \label{tab:parameters range}
  \end{center}
\end{table}

\begin{figure}
    \begin{subfigure}{0.5\textwidth}
    \centering
    \caption{}
    \includegraphics[width=0.9\textwidth]{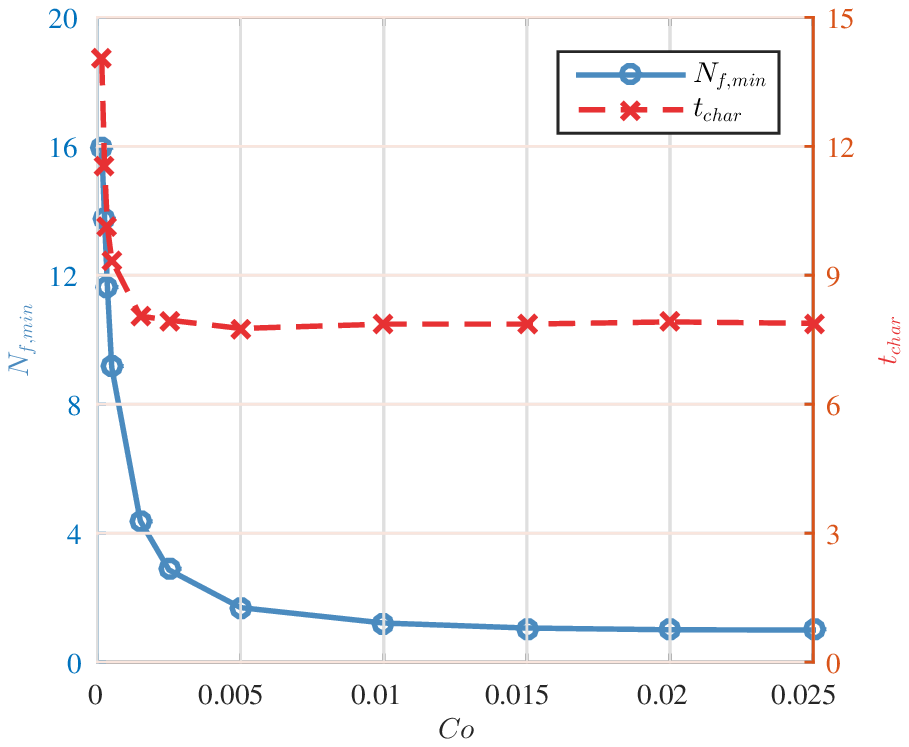}
    \label{fig:dependence of flocculation Co}
    \end{subfigure}
    \begin{subfigure}{0.5\textwidth}
    \centering
    \caption{}
    \includegraphics[width=0.9\textwidth]{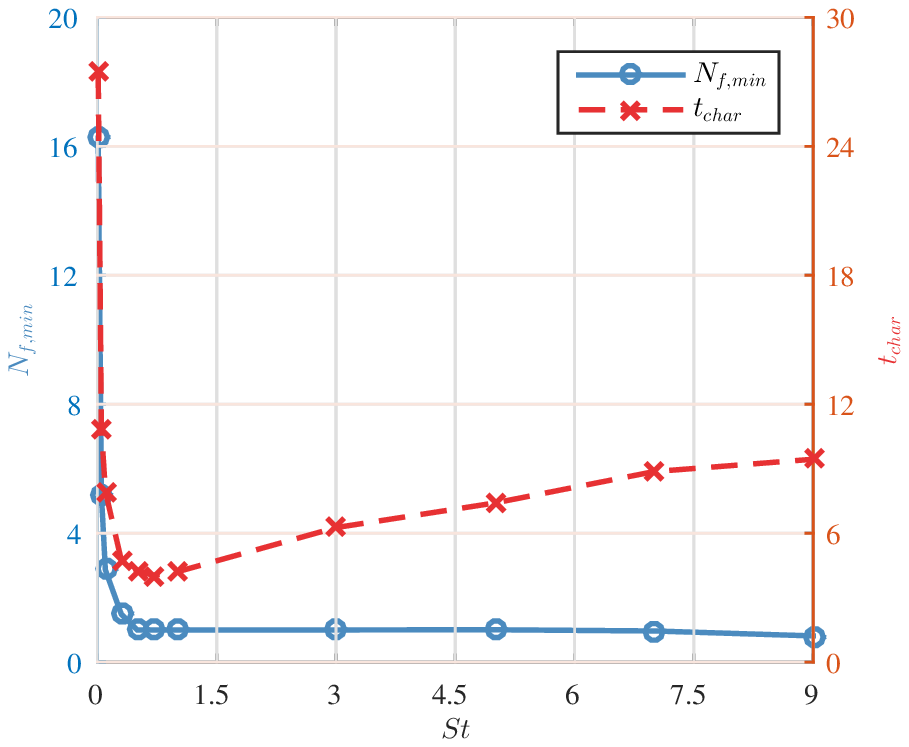}
    \label{fig:dependence of flocculation St} 
    \end{subfigure}
    \caption{The flocculation time scale and the equilibrium number of flocs as functions of $Co$ and $St$, respectively, for $N_p = 50$, $D_p = 0.1$, $\rho_s = 1$ and $W = 0$: (a) Influence of the cohesive number $Co$, for $St = 0.1$; (b) Influence of the Stokes number $St$, for $Co = 2.5 \times 10^{-3}$.}
    \label{fig:dependence of flocculation}
\end{figure}

In order to explore the dependence of the flocculation process on the key governing quantities, we carry out 
\bernhard{a total of 300}
simulations covering the parameter ranges listed in table \ref{tab:parameters range}. Figure \ref{fig:dependence of flocculation Co} shows that the number of flocs during the equilibrium stage $N_{f,min}$ decreases for increasing $Co$. Beyond $Co \approx 0.01$, all of the primary particles aggregate into one large floc, as the cohesive forces overwhelm the hydrodynamic stresses trying to break up the floc. The characteristic flocculation time initially decreases as $Co$ grows, and then levels off and remains constant. Figure \ref{fig:dependence of flocculation St} indicates that for large Stokes numbers $St$ the equilibrium floc number $N_{f,min}$ also asymptotically approaches one. Interestingly, we observe that the flocculation time $t_{char}$ displays a pronounced minimum around $St \approx 0.7$, which reflects the well-known optimal coupling between particle and fluid motion when the particle response time is of the same order as the characteristic timescale of the flow \citep{Wang1993}. Under these conditions, particles rapidly accumulate near the edges of the vortices, which facilitates the formation of flocs.

\subsection{A new flocculation model based on the simulation data} \label{subsec:A new model for flocculation process}

\begin{figure}
    \centering
    \includegraphics[width=0.9\textwidth]{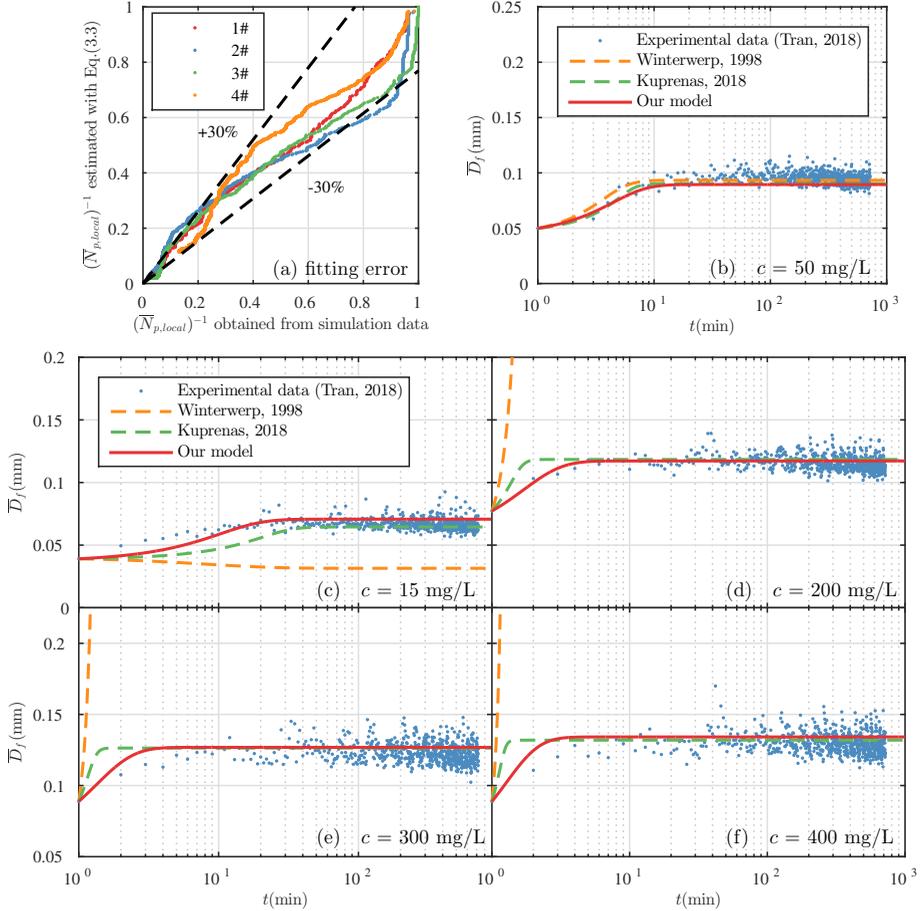}
    \caption{\Rthree{(a) Total fitting deviation of eqns. (\ref{eq:Our new model Np,local})-(\ref{eq:Our new model b}), cases $1\#\sim4\#$ shown here have the largest fitting error among 300 different tests for the cellular flow field (figure \ref{fig:streamlines}).} \aleck{Comparison with experimental data: (b) calibration of the empirical coefficients for the models of \cite{Winterwerp1998floc} ($k_A ^{'} = 1.35$ and $k_B ^{'} = 1.29 \times 10^{-5}$), \cite{Kuprenas2018floc} ($k_A ^{'} = 0.45$ and $k_B ^{'} = 1.16 \times 10^{-6}$), and for our eqns. (\ref{eq:Our new model Np,local,max})-(\ref{eq:Our new model b}) ($a_1 = 500$ and $a_2 = 35$);} (c)-(f) Comparison between experimental data and predictions by \Rone{the model of \citet{Kuprenas2018floc} and 
    eqns. (\ref{eq:Our new model Df})-(\ref{eq:Our new model b}), the predictions by \citet{Winterwerp1998floc} are also shown for further comparison.}
    The experimental parameters are $D_p=5 \rm{\mu m}$, $G = 50 \rm{s^{-1}}$, $\rho_p=2650 \rm{kg/m^3}$, $\rho_f=1000 \rm{kg/m^3}$, $\mu_f = 0.001 \rm{Ns/m^2}$, $c=15 \sim 400 \rm{mg/L}$.}
    \label{fig:verification and comparison}
\end{figure}

According to \citet{Khelifa2006II, Khelifa2006I}, for flocs of fractal dimension $n_f$ the mean floc size $\overline D_f$ is related to the average number of 
\Rone{primary particles per floc $\overline N_{p,local}=N_p/N_f$:}
\begin{equation}\label{eq:Our new model Df}
         \overline D_f = (\overline N_{p,local}) ^ {\frac{1}{n_f}} D_p \ .
\end{equation}
Eqn. (\ref{eq:exponential fitting}) yields for the average number of particles per floc 
\begin{equation} \label{eq:Our new model Np,local}
         \overline N_{p,local} = \frac{1}{(1 / \overline N_{p,local,int} - 1 / \overline N_{p,local,max}) e^{b t} +  1 / \overline N_{p,local,max}} \ ,
\end{equation}
where the initial number of particles per floc is $\overline N_{p,local,int} = N_p / N_{f,int}$, and the average number of particles per floc during the equilibrium stage is $\overline N_{p,local,max} = N_p / N_{f,min}$. Fitting the simulation results over the parameter ranges listed in table \ref{tab:parameters range} yields
\begin{subequations} \label{eq:Our new model Np,local,max}
    \begin{eqnarray} \label{eq:Our new model Np,local,max a}
            \overline N_{p,local,max} & = & 8.5 a_1 St^{0.65} Co^{0.58} D_p^{-2.9} \phi^{0.39} \rho_s^{-0.49} (W+1)^{-0.38} \ , \\
            \overline N_{p,local,max} & = & N_p,  \quad  if \ \overline N_{p,local,max} \geqslant N_p \ , 
    \end{eqnarray}
\end{subequations}
\begin{equation} \label{eq:Our new model b}
    b = \left\{
    \begin{array}{ll}
        -0.7 a_2 St^{0.36} Co^{-0.017} D_p^{-0.36} \phi^{0.75} \rho_s^{-0.11} (W+1)^{-1.4},          &  St \leqslant 0.7  \ , \\ 
        -0.3 a_2 St^{-0.38} Co^{0.0022} D_p^{-0.61} \phi^{0.67} \rho_s^{0.033} (W+1)^{-0.46},                 &  St > 0.7  \ .
    \end{array}\right.
\end{equation}
For the present cellular model flow the values $a_1 = a_2 = 1$ in eqns. (\ref{eq:Our new model Np,local,max a}) and (\ref{eq:Our new model b}) yield optimal agreement with the simulation data \Rthree{with the fitting deviation of $\pm 30\%$ (figure \ref{fig:verification and comparison}a)}. For real turbulent flows, we will determine $a_1$ and $a_2$ by calibrating with experimental data, as will be explained below.

\citet{Winterwerp1998floc} introduced a population balance model that accounts for aggregation and breakage 
\Rone{for low turbulence levels.}
His model has the form
\begin{equation}\label{eq:Winterwerp1998}
         \frac{\mathrm{d} \overline D_f}{\mathrm{d} t} = \frac{k_A ^{'}}{n_f} \frac{D_p^{n_f-3}}{\rho_p} G c \overline D_f^{4-n_f} - \frac{k_B ^{'}}{n_f} (\frac{\mu_f}{F_y})^q (\frac{\overline D_f - D_p}{D_p})^p G^{q+1} \overline D_f^{2q+1} \ ,
\end{equation}
where $G$ indicates the shear rate of the turbulence (units $\rm s^{-1}$), $c$ represents the sediment concentration ($\rm kg/m^3$), and $F_y$ denotes the yield strength of the flocs ($\rm N$). Winterwerp suggests the values $F_y = 10^{-10} \rm N$, $n_f = 2$, $p = 1$ and $q = 0.5$. The empirical coefficients $k_A ^{'}$ and $k_B ^{'}$ depend on the physico-chemical properties of the sediment and fluid. While this model has enjoyed wide popularity in the literature \citep{Winterwerp2006floc, Son2009floc, Son2008floc, Lee2011, Keyvani2014floc, Strom2016floc}, it has also been pointed out that for large turbulent shear and sediment concentrations the model predicts that the floc size will be larger than the Kolmogorov scale $\eta$, which is not consistent with experimental observations \citep{Keyvani2014floc, Kuprenas2018floc, Sherwood2018floc}. To address this issue, \citet{Kuprenas2018floc} recently suggested the modification $q = 0.5 + 1.5 \overline D_f / \eta$. In the following, we will compare predictions by the current model with both of these earlier models.

\citet{Tran2018experiment} measured the floc size $\overline D_f(t)$ in turbulence for constant shear rate $G$ and sediment concentration $c$ (figure \ref{fig:verification and comparison}b-\ref{fig:verification and comparison}f). \aleck{They determined the empirical coefficients $k_A ^{'}$ and $k_B ^{'}$ by calibrating with the experimental data for $c = 50 \rm{mg/L}$. In a similar fashion, we will determine the constants $a_1$ and $a_2$ required for our model by calibrating with the same case displayed in figure \ref{fig:verification and comparison}b.} Towards this end, we need to convert the experimental data into characteristic length $L$ and velocity $U_0$ scales that can be employed in our model, eqns. (\ref{eq:Our new model Np,local})-(\ref{eq:Our new model b}). We accomplish this by setting $L = \eta = [\mu_f/ (\rho_f G)]^{0.5}$ and $U_0 = G \eta/4$. Furthermore, we assume the Hamaker constant to be $A_H = 1.0 \times 10^{-20} \rm J$ \citep[see][p. 37-39]{Vowinckel2018Cohesive}, and the fractal dimension $n_f = 2$, \aleck{which yields the correction constants $a_1 = 500$ and $a_2 = 35$.} The mean floc size can then be obtained from eqn. (\ref{eq:Our new model Df}). For lower sediment loadings, figure \ref{fig:verification and comparison}c shows that our model yields predictions similar to those of Kuprenas for the equilibrium floc size, while it tends to perform somewhat better than Kuprenas' model during the transient stage. These differences become more pronounced for higher sediment loadings (figure \ref{fig:verification and comparison}d-\ref{fig:verification and comparison}f), and they likely reflect the more realistic modeling of the particle-particle interactions in the present simulations. \Rone{For these particular flow conditions, Winterwerp's model yields valid predictions only for sediment loads below approximately 200mg/L}. \Rtwo{Hence, the results encourage the use of our conceptually simplified cellular flow model as a cost-efficient tool to derive scaling laws in the form of eqns. (\ref{eq:Our new model Df})-(\ref{eq:Our new model b}) for a wide parameter range for flocculation of cohesive particles in turbulent flow conditions.}

\section{Conclusions} \label{sec:Conclusions}

We have analyzed the flocculation dynamics of cohesive sediment via one-way coupled simulations in a model turbulent flow field. The computational model accounts for Stokes drag, lubrication, cohesive and direct contact forces, and it yields the time-dependent floc size as a function of the governing dimensionless parameters. The simulations reproduce the transient growth of the cohesive flocs, as well as the emergence of a log-normal equilibrium distribution governed by the balance of aggregation and breakage. By accounting for the detailed physical mechanisms governing particle-particle interactions, the simulations demonstrate that flocculation proceeds most rapidly when the Stokes number of the primary particles is \textit{O}(1). We employ the computational data in order to propose a new flocculation model. As it is based on a more realistic representation of particle-particle interactions, this new model yields improved agreement with the experimental measurements of \citet{Tran2018experiment}, especially during the transient stages.

\vspace{.2in}
\noindent
{\bf Acknowledgements}\\
The authors thank D. Tran for providing his experimental data for comparison purposes. EM gratefully acknowledges support through NSF grants CBET-1803380 and OCE-1924655, as well as by the Army Research Office through grant W911NF-18-1-0379. TJH received support through NSF grant OCE-1924532. KZ is supported by the China Scholarship Council, as well as by the China National Fund for Distinguished Young Scientists through grant 51425603. BV gratefully acknowledges support through German Research Foundation (DFG) grant VO2413/2-1. Computational resources for this work used the Extreme Science and Engineering Discovery Environment (XSEDE), which is supported by NSF grant TG-CTS150053.

\bibliographystyle{jfm}
\bibliography{jfm-instructions}

\end{document}